\documentclass[pdftex, pra, twocolumn, floatfix, superscriptaddress, nofootinbib]{revtex4}

\usepackage{times}
\usepackage{amsfonts}
\usepackage{amssymb}
\usepackage{amsmath}
\usepackage{graphicx}
\usepackage{subfigure}
\usepackage[colorlinks=true, breaklinks=true, linkcolor=blue, citecolor=blue, urlcolor=blue]{hyperref}
\usepackage{bbm}

\newcommand{\doi}[2]{\href{https://doi.org/#1}{#2}}
\newcommand{\arxiv}[1]{\href{https://arxiv.org/abs/#1}{arXiv:#1}}

\begin{document}

\title{Tuning an effective spin chain of three strongly interacting one-dimensional \\ fermions with the transversal confinement}

\author{Frank Deuretzbacher}
\email{frank.deuretzbacher@itp.uni-hannover.de}
\affiliation{Institut f\"ur Theoretische Physik, Leibniz Universit\"at Hannover, Appelstrasse 2, DE-30167 Hannover, Germany}

\author{Luis Santos}
\affiliation{Institut f\"ur Theoretische Physik, Leibniz Universit\"at Hannover, Appelstrasse 2, DE-30167 Hannover, Germany}

\begin{abstract}
Strongly interacting one-dimensional fermions form an effective spin chain in the absence of an external lattice potential. We show that the exchange coefficients of such a chain may be locally tuned by properly tailoring the transversal confinement. In particular, in the vicinity of a confinement-induced resonance~(CIR) the exchange coefficients may have simultaneously opposite ferromagnetic and antiferromagnetic characters at different locations along the trap axis. Moreover, the local exchanges may be engineered to induce avoided crossings between spin states at the CIR, and hence a ramp across the resonance may be employed to create different spin states and to induce spin dynamics in the chain. We show that such unusual spin chains have already been realized in the experiment of Murmann {\it et al.} [\href{https://doi.org/10.1103/PhysRevLett.115.215301}{Phys. Rev. Lett. {\bf 115}, 215301 (2015)}].
\end{abstract}

\maketitle

\section{Introduction}

Ultracold atoms are particularly well suited to study strongly interacting one-dimensional~(1D) Tonks-Girardeau gases due to the high degree of control and tunability of these systems~\cite{Kinoshita04, Kinoshita06, Haller09, Paredes04}. This motivated the generalization of the single-component Tonks-Girardeau gas~\cite{Girardeau60} to multicomponent systems~\cite{Girardeau07, Deuretzbacher08, Guan09} and the development of a perturbation theory for the strongly-interacting regime~\cite{Volosniev14}, which is also applicable to attractive interactions~\cite{Astrakharchik05}. Moreover, it was found that strongly interacting multicomponent systems form tunable spin chains~\cite{Deuretzbacher14}, which allow for the simulation of quantum magnetism in the absence of a lattice~\cite{Deuretzbacher14, Volosniev15, Yang15, Massignan15, Yang16a, Yang16b, Hu16a, Yang16c, Hu16b, Deuretzbacher17, Dehkharghani17}. Such spin chains were recently realized experimentally \cite{Murmann15a, Zuern12a, Pagano14, Meinert16}.

In this paper, we show that the exchange coefficients of these spin chains may be locally tuned by varying the strength of the transversal confinement along the weak axis of the quasi-1D trap. In particular, in the vicinity of a confinement-induced resonance~(CIR) it is possible to engineer spin chains with simultaneously ferromagnetic~(FM) and antiferromagnetic~(AFM) exchange coefficients. This local control of the spin chain opens interesting possibilities for the preparation of few-body spin states, and for the controlled study of spin dynamics in the effective chain. Interestingly, we show that such unusual spin chains have been already experimentally realized in Ref.~\cite{Murmann15a}. In particular, we demonstrate that the inhomogeneous transversal confinement resulted in spatially inhomogeneous spin exchanges that explain up to now unclear features observed in those experiments in the vicinity of the CIR.

The structure of the paper is as follows. Section~\ref{sec:tunneling-experiments} briefly discusses the experiments of Ref.~\cite{Murmann15a}. Section~\ref{sec:spin-chain-model} introduces the spin-chain model, whereas Sec.~\ref{sec:tunneling-model} is devoted to the outcoupling theory. Section~\ref{sec:invg} discusses the spatially varying interaction strength at the focus of the Gaussian laser beam that creates the quasi-1D trap. The explanation of the anomalous outcoupling results observed at the CIR in Ref.~\cite{Murmann15a} is discussed in Sec.~\ref{sec:origin-of-the-peak}. In Sec.~\ref{sec:unusual-J} we illustrate with a simple example the engineering of spin chains using the transversal confinement. Finally, Sec. 
\ref{sec:conclusions} summarizes our conclusions.

\begin{figure}
\begin{center}
\includegraphics[width = \columnwidth]{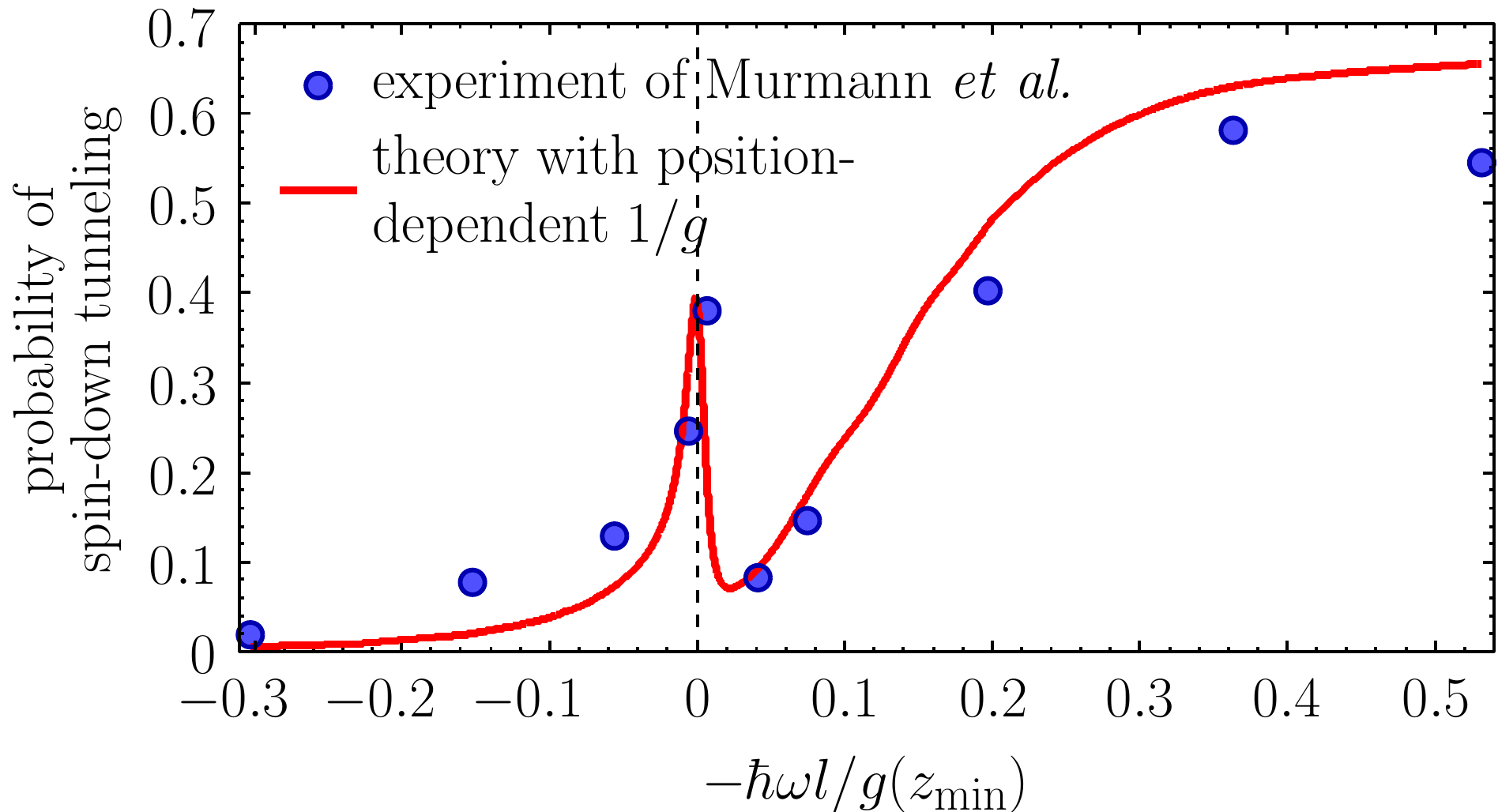}
\caption{Tunneling probability of the spin-down atom for a spin chain consisting of two spin-up and one spin-down atom with antiferromagnetic~(AFM) spin order as a function of the inverse interaction strength at the trap minimum, $-1/g(z_\mathrm{min})$. $\omega$ and $l$ are the frequency and length scale of the harmonic oscillator.}
\label{fig:tunneling}
\end{center}
\end{figure}

\section{Tunneling experiments}
\label{sec:tunneling-experiments}

\begin{figure}
\begin{center}
\includegraphics[width = 0.8 \columnwidth]{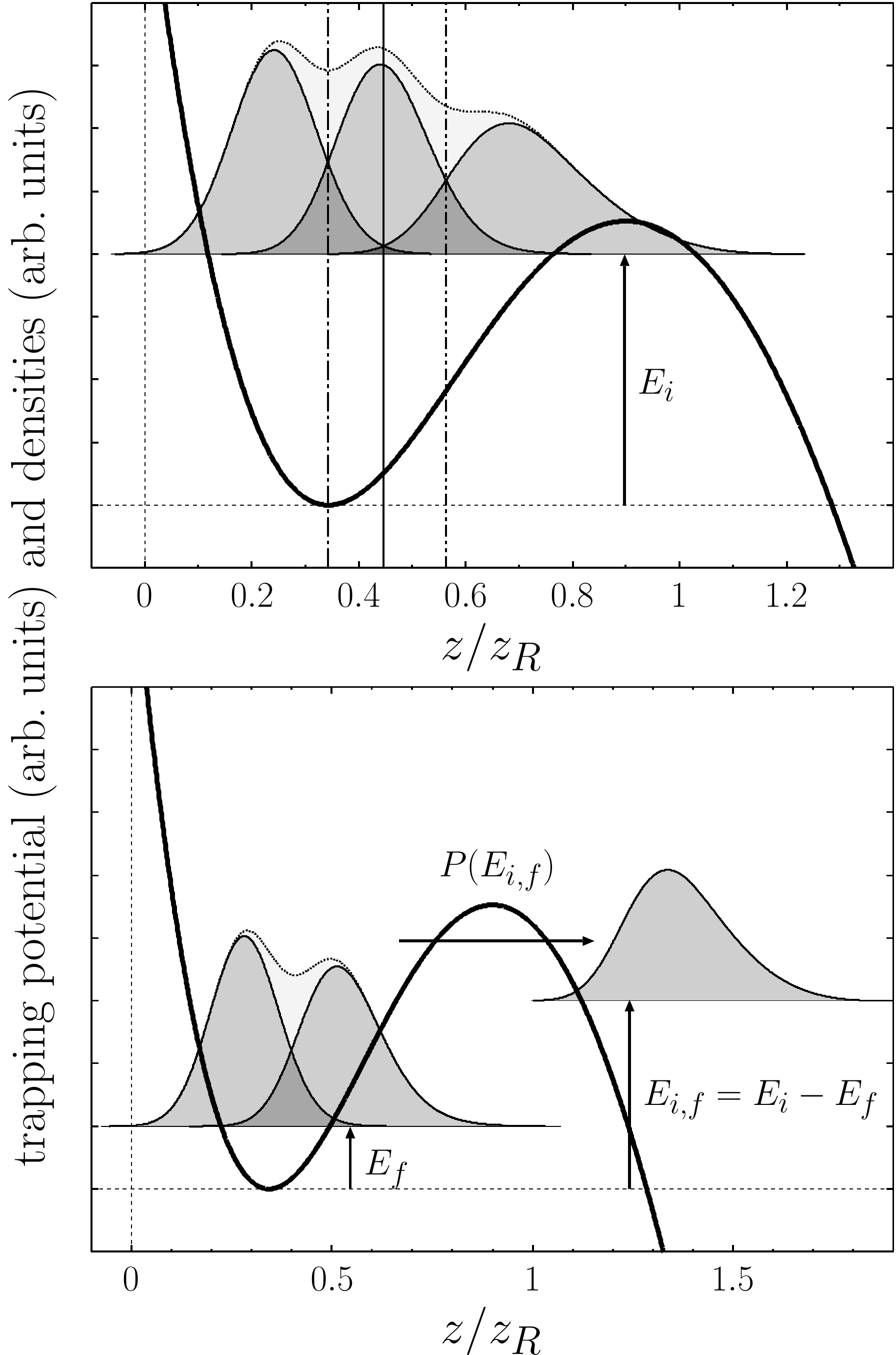}
\caption{(Top) Atom chain with energy $E_i$ before tunneling of the rightmost atom. Vertical lines mark the position~$z=0$, the position of maximum overlap between the first and second atom, the mean position of the second atom, and the position of maximum overlap between the second and third atom (from left to right). (Bottom) Atom chain after tunneling of the rightmost atom. The two atoms, which stay in the trap, have energy $E_f$. The tunneling atom has energy $E_{i,f} = E_i - E_f$. The atom tunnels with the energy-dependent probability $P(E_{i,f})$ through the barrier. $z_R$ is the Rayleigh length.}
\label{fig:setup}
\end{center}
\end{figure}

The experiments in Ref.~\cite{Murmann15a} studied the spatial spin distribution of quasi-1D two-component fermions $\{\uparrow,\downarrow\}$ considering different few-body admixtures with an exquisite control of the number of atoms $N_\uparrow$ and $N_\downarrow$ in each component. We focus below for simplicity and for its experimental relevance, on the particular case of $N_\uparrow = 2$ and $N_\downarrow = 1$, but similar arguments may be employed for other cases. In Ref.~\cite{Murmann15a} an AFM spin order was realized in a quasi-1D trap by first preparing the noninteracting ground state in which the atoms occupied the lowest trap levels according to the Pauli exclusion principle. Subsequently, the effective 1D interaction strength $g$ was ramped by means of a magnetic-field Feshbach resonance up to a large final value in the strongly interacting regime around the CIR. The spatial spin distribution in the trap was studied by means of a tunneling technique: The trap was tilted such that the rightmost atom could tunnel out~(Fig.~\ref{fig:setup}), and the spin of the outcoupled atom was measured. The probability for a spin down was measured as a function of $1/g$ and related to the spin distribution in the trap.

The spin-down probability of the outcoupled atom presented a marked peak at the CIR~(Fig.~\ref{fig:tunneling}) whose origin remained unclear. We show in Sec.~\ref{sec:invg} that the focused Gaussian beam that generates the quasi-1D trap induces an inhomogeneous transversal confinement along the trap axis~\cite{Gharashi15}, that in turn results in a spatially inhomogeneous interaction strength $g$. As discussed below, the latter has a strong influence on the spin configurations of the spin chain in the immediate vicinity of a CIR that explain the observed peak in Ref.~\cite{Murmann15a}.

\section{Effective spin chain}
\label{sec:spin-chain-model}

The physics of quasi-1D multicomponent atoms in the strongly interacting regime may be well understood by means of an effective spin-chain model. Indeed, in the strongly interacting regime (in the vicinity of a CIR), the atoms order in a row. The atomic density is then that of spinless fermions, whereas the spin degree of freedom is governed by spin exchanges as those of a spin chain, but in the absence of any external lattice potential~\cite{Deuretzbacher14}. In the case of spin-$\frac{1}{2}$ fermions considered here, the effective spin-chain Hamiltonian is given by a Heisenberg model,
\begin{equation} \label{eq:Hs}
H = \left[ E_F^{(N)} - \frac{1}{2} \sum_{i=1}^{N-1} J_i \right] \openone + \frac{1}{2} \sum_{i=1}^{N-1} J_i \, \vec \sigma^{(i)} \cdot \vec \sigma^{(i+1)} ,
\end{equation}
where $\vec \sigma$ is the Pauli vector and $E_F^{(N)}$ is the energy of $N$ spinless noninteracting fermions in the 1D trap $V(z)$. The exchange coefficients $J_i$ depend on the position within the spin chain and are given by \cite{Volosniev14, Deuretzbacher14}
\begin{equation}
J_i = \frac{N! \hbar^4}{m^2} \int dz \, \frac{C_i(z)}{g(z)} ,
\end{equation}
with
\begin{eqnarray}
C_i(z_{i+1}) & = & \int dz_1 \dotsi dz_i dz_{i+2} \dotsi dz_N \, \delta(z_i-z_{i+1}) \nonumber \\
& & \times \theta (z_1, \dotsc, z_N) \left| \frac{\partial \psi_F}{\partial z_i} \right|^2 \! .
\end{eqnarray}
Here, $\theta (z_1, \dotsc , z_N) = 1$ if $z_1 < \dotsb < z_N$, and zero otherwise, and $\psi_F$ is the ground-state wave function of $N$ spinless noninteracting fermions in the 1D trapping potential $V(z)$. In a good approximation, the exchange coefficients are also proportional to the local density cubed and to the local inverse interaction strength, $J_i \propto n^3/g$ \cite{Deuretzbacher14, Yang16b}.

For the particular case of ${N_\uparrow = 2}$ and ${N_\downarrow = 1}$, the eigenstates in a symmetric trap (${J_1 = J_2}$) are~\cite{Murmann15a} an AFM state,
\begin{equation}
| \mathrm{AFM} \rangle = \frac{1}{\sqrt{6}} \Bigl( | \uparrow , \uparrow , \downarrow \rangle - 2 | \uparrow , \downarrow , \uparrow \rangle + | \downarrow , \uparrow , \uparrow \rangle \Bigr) ,
\end{equation}
an intermediate~(IM) state,
\begin{equation}
| \mathrm{IM} \rangle = \frac{1}{\sqrt{2}} \Bigl( | \uparrow , \uparrow , \downarrow \rangle - | \downarrow , \uparrow , \uparrow \rangle \Bigr) ,
\end{equation}
and an FM state,
\begin{equation}
| \mathrm{FM} \rangle = \frac{1}{\sqrt{3}} \Bigl( | \uparrow , \uparrow , \downarrow \rangle + | \uparrow , \downarrow , \uparrow \rangle + | \downarrow , \uparrow , \uparrow \rangle \Bigr) ,
\end{equation}
where the spin states, e.g., $| \uparrow , \uparrow , \downarrow \rangle$, denote the spatial distribution from left to right of the three spins.

In the ${\{ | \mathrm{AFM} \rangle}$, ${| \mathrm{IM} \rangle}$, ${| \mathrm{FM} \rangle \}}$ basis, the spin Hamiltonian~(\ref{eq:Hs}) takes the form
\begin{equation}
H = E_F^{(3)} \openone +
\begin{bmatrix}
  -\frac{3}{2}(J_1+J_2) & \frac{\sqrt{3}}{2}(J_1-J_2) & 0 \\
  \frac{\sqrt{3}}{2}(J_1-J_2) & -\frac{1}{2}(J_1+J_2) & 0 \\
  0 & 0 & 0
\end{bmatrix} \! .
\end{equation}
An imbalance of $J_1$ and $J_2$ therefore couples the states ${| \mathrm{AFM} \rangle}$ and ${| \mathrm{IM} \rangle}$ of the symmetric trap. The resulting states of the imbalanced system ($J_1 \neq J_2$) are thus given by
\begin{equation}
| \mathrm{AFM}' \rangle = \cos \frac{\varphi}{2} | \mathrm{AFM} \rangle - \sin \frac{\varphi}{2} | \mathrm{IM} \rangle ,
\end{equation}
\begin{equation}
| \mathrm{IM}' \rangle = \cos \frac{\varphi}{2} | \mathrm{IM} \rangle + \sin \frac{\varphi}{2} | \mathrm{AFM} \rangle ,
\end{equation}
and $| \mathrm{FM}' \rangle = | \mathrm{FM} \rangle$ with
\begin{equation} \label{eq:angle}
\varphi = \arctan \left( \sqrt{3} \frac{J_1-J_2}{J_1+J_2} \right) \! .
\end{equation}
In the experiment \cite{Murmann15a}, the system was prepared in the AFM state of the tilted trap, ${| \mathrm{AFM}' \rangle}$.

\section{Outcoupling}
\label{sec:tunneling-model}

The results of tunneling experiments as those described in Sec.~\ref{sec:tunneling-experiments} may be well understood from the spatial spin distribution. In the regime of nearly infinite repulsion between the atoms, only the rightmost atom of the spin chain can tunnel through the barrier of the tilted trap, since the order of the particles is fixed (see~Fig.~\ref{fig:setup}). The probability of spin-down tunneling is therefore the probability of the rightmost spin of the $| \mathrm{AFM}' \rangle$ state to be $\downarrow$,
\begin{equation} \label{eq:spin-orientation}
\mspace{-6mu} P_\downarrow \approx |\langle \uparrow , \uparrow , \downarrow | \mathrm{AFM}' \rangle|^2 = \left( \cos \frac{\varphi}{2} \frac{1}{\sqrt{6}} - \sin \frac{\varphi}{2} \frac{1}{\sqrt{2}} \right)^{\!2} \! .
\end{equation}
This simple formula is a good approximation for $P_\downarrow$ if $|\hbar \omega l / g| \lesssim 0.1$ ($\omega$ and $l$ are the frequency and length scale of the harmonic oscillator).

Outside this regime, $P_\downarrow$ depends also on the energies of the triplet and singlet states, $E_F^{(2)}$ and $E_F^{(2)} - 2 J_1$, in which the remaining two spins are after the tunneling process, since the probability that the rightmost atom tunnels through the barrier depends on the energy differences, $E_{\mathrm{AFM}'} - E_F^{(2)}$ and $E_{\mathrm{AFM}'} - \bigl( E_F^{(2)} - 2 J_1 \bigr)$, which are transferred to the tunneling atom. 
More precisely, if the tunneling atom has spin down, the final state of the system is ${| f_1 \rangle = | \uparrow , \uparrow \rangle \otimes | \downarrow \rangle}$, while if the tunneling atom has spin up, the remaining two atoms are either in the triplet or singlet state, i.e., the final states are ${| f_2 \rangle = \frac{1}{\sqrt{2}} \left( | \uparrow , \downarrow \rangle + | \downarrow , \uparrow \rangle \right) \otimes | \uparrow \rangle}$ or ${| f_3 \rangle = \frac{1}{\sqrt{2}} \left( | \uparrow , \downarrow \rangle - | \downarrow , \uparrow \rangle \right) \otimes | \uparrow \rangle}$. The rate of a transition between the initial state $| i \rangle = | \mathrm{AFM}' \rangle$ to one of the final states $| f \rangle$ is proportional to the squared spin overlap $|\langle i | f \rangle|^2$ and to the probability $P(E_{i,f}) = E_{i,f} e^{-2 \gamma (E_{i,f})}$ that an atom with energy $E_{i,f}$ tunnels out of the tilted trap,
\begin{equation}
T_{i,f} \propto |\langle i | f \rangle|^2 E_{i,f} e^{-2 \gamma (E_{i,f})} ,
\end{equation}
with $E_{i,f} = E_i - E_f$ and
\begin{equation}
\gamma(E_{i,f}) = \frac{1}{\hbar} \int_{z_1}^{z_2} dz \, \sqrt{2 m \bigl[ V(z) - E_{i,f} \bigr]} ,
\end{equation}
where $z_1$ and $z_2$ are the intersection points of $E_{i,f}$ with the barrier. The probability of spin-down tunneling is now
\begin{equation}
P_\downarrow = \frac{T_{i,f_1}}{T_{i,f_1} + T_{i,f_2} + T_{i,f_3}} .
\end{equation}
For a positive $J_1$, tunneling into the singlet state $| f_3 \rangle$ is energetically favored, and therefore $P_\downarrow \approx 0$ for $-\hbar \omega l / g \lesssim -0.1$, while for a negative $J_1$, tunneling into the two triplet states $| f_1 \rangle$ and $| f_2 \rangle$ is energetically favored, and therefore $P_\downarrow \approx 2/3$ for $-\hbar \omega l / g \gtrsim 0.3$, since $|\langle i | f_1 \rangle|^2 = 2 |\langle i | f_2 \rangle|^2$.

\section{Position-dependent interaction strength}
\label{sec:invg}

The quasi-1D optical dipole trap is created at the focus of a single far red-detuned Gaussian laser beam. The intensity distribution of the laser beam is given by \cite{Grimm00, Zuern12b, Murmann15b}
\begin{equation}
I \propto \frac{e^{-2 \rho^2 / w^2(z)}}{w^2(z)} ,
\end{equation}
with $w(z) = w_0 \sqrt{1 + (z/z_R)^2}$, with $z_R = \pi w_0^2 / \lambda$ the Rayleigh length. In the experiments of Ref.~\cite{Murmann15a}, the minimal beam waist is $w_0 = 1.838 \, \mu\mathrm{m}$ and the wavelength is $\lambda = 1064 \, \mathrm{nm}$. The potential experienced by the atoms is proportional to this intensity distribution and given by
\begin{equation}
V = - \frac{p V_0}{1 + \left( z / z_R \right)^2} \exp \mspace{-5mu} \left[ - \frac{2 \rho^2}{w_0^2 \bigl( 1 + \left( z / z_R \right)^2 \bigr)} \right] \! ,
\end{equation}
where $p$ is the trap depth, and $V_0$ is the initial depth of the optical potential~($p \approx 0.7-1.4$ and $V_0/k_B\simeq 3.326~\mu\mathrm{K}$ in Ref.~\cite{Murmann15a}). A Taylor expansion of the Gaussian yields
\begin{equation}
V = p V_0 \left[ - \frac{1}{1 + \left( z / z_R \right)^2} + \frac{2 \rho^2}{w_0^2 \bigl( 1 + \left( z / z_R \right)^2 \bigr)^{\!2}} \right] \! .
\end{equation}
The first term generates the axial potential and the second term generates a radial harmonic confinement, which becomes weaker with increasing $\left( z / z_R \right)^2$. Therefore, we obtain a position-dependent transverse trap frequency \cite{Gharashi15},
\begin{equation} \label{eq:omega_perp}
\omega_\perp(z) = \omega_\perp^* \frac{\sqrt{p}}{1 + \left( z / z_R \right)^2} ,
\end{equation}
where $\omega_\perp^* = \sqrt{4 V_0 / (m w_0^2)}$ is the transverse trap frequency for $p=1$ and $z=0$. Accordingly, the transverse oscillator length becomes position dependent,
\begin{equation} \label{eq:l_perp}
l_\perp(z) = l_\perp^* \frac{\sqrt{1 + \left( z / z_R \right)^2}}{p^{1/4}} ,
\end{equation}
with $l_\perp^* = \sqrt{\hbar / (m \omega_\perp^*)}$. Additionally, in the experiment of Ref.~\cite{Murmann15a}, a linear magnetic-field gradient was applied along the $z$ axis, $V_\mathrm{mag} = -\mu_m B' z$, to allow for the tunneling of atoms out of the trap. Here, $\mu_m$ is the magnetic moment of the atoms and $B'$ is the strength of the gradient (${B' \simeq 18.92 \, \mathrm{G} / \mathrm{cm}}$ in Ref.~\cite{Murmann15a}).

In a quasi-1D trap with a strong homogeneous transversal harmonic confinement of frequency $\omega_\perp$, the strength of the contact interaction $g$ is given by~\cite{Olshanii98}
\begin{equation} \label{eq:Olshanii}
2 \hbar \omega_\perp l_\perp / g = l_\perp / a_\mathrm{3D} - C
\end{equation}
with the bare three-dimensional (3D) scattering length $a_\mathrm{3D}$ and $C \simeq 1.0326 \dotsc$. Since $\omega_\perp(z)$ and $l_\perp(z)$ vary slowly on the length scale of the axial trap, we may insert~(\ref{eq:omega_perp}) and~(\ref{eq:l_perp}) into Eq.~(\ref{eq:Olshanii}) to obtain a position-dependent inverse interaction strength,
\begin{equation} \label{eq:invg}
\frac{\hbar \omega_\perp^* l_\perp^*}{g(z)} = \frac{\sqrt{1 + \left( z / z_R \right)^2}}{2 p^{1/4}} \! \left[ \frac{l_\perp^* \sqrt{1 + \left( z / z_R \right)^2}}{p^{1/4} a_\mathrm{3D}} - C \right] \! .
\end{equation}
This position dependence is particularly important in the vicinity of the CIR, since $1/g(z)$ may then have a zero crossing within the trap. As shown below, this explains ultimately the unclear features observed in the tunneling experiments of Ref.~\cite{Murmann15a}.

\begin{figure}
\begin{center}
\includegraphics[width = 0.8 \columnwidth]{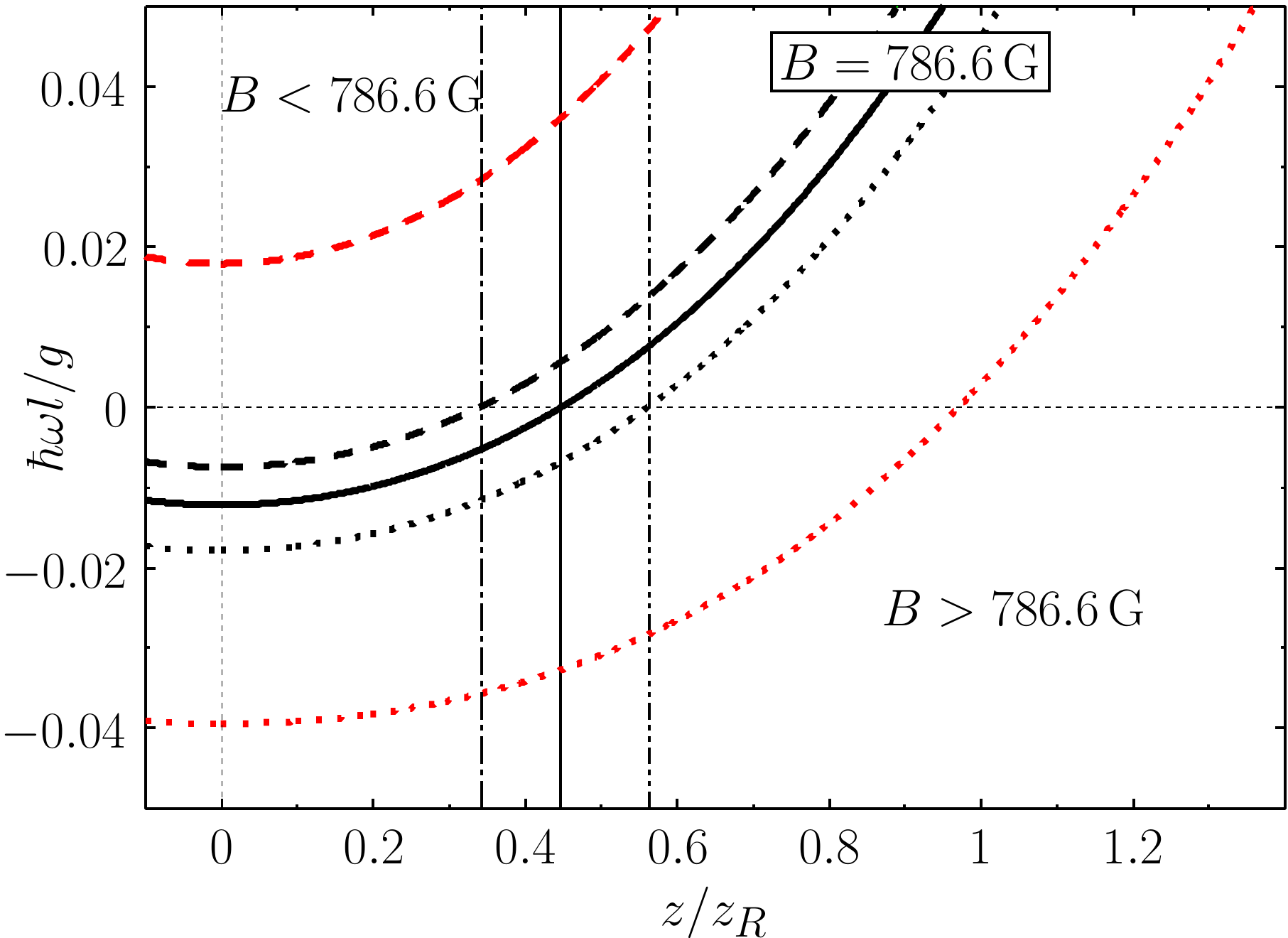}
\caption{Inverse interaction strength $1/g$ as a function of the position in the trap $z$ for different $B$-field strengths around the CIR. The solid line is for $B = 786.6\,\mathrm{G}$, the dashed lines are for $B < 786.6\,\mathrm{G}$, and the dotted lines are for $B > 786.6\,\mathrm{G}$. The thin vertical lines mark the same positions as in Fig.~\ref{fig:setup} (top). $\omega$ and $l$ are the oscillator frequency and length and $z_R$ is the Rayleigh length.}
\label{fig:invg}
\end{center}
\end{figure}

\section{Anomalous tunneling at the CIR}
\label{sec:origin-of-the-peak}

The anomalous behavior of the outcoupling probability $P_\downarrow$ at the CIR~(Fig.~\ref{fig:tunneling}) results from the strong variation of the ratio $J_1/J_2$ in the vicinity of the CIR, which determines the orientation of the rightmost spin of the spin chain [see Eqs.~(\ref{eq:angle}) and (\ref{eq:spin-orientation})]. As discussed in Sec.~\ref{sec:invg}, the weakening of the radial harmonic confinement with increasing axial distance from the focus of the optical trap $|z/z_R|$ leads to a spatial dependence of the inverse effective 1D interaction strength $1/g$ given by Eq.~(\ref{eq:invg}). Moreover, the center of the tilted trap is displaced from the focus of the laser beam, i.e., $z_\mathrm{min} > 0$ (see Fig.~\ref{fig:setup}). As a result, $1/g$ is smaller between the first and second atom than between the second and third atom (see the thick curves in Fig.~\ref{fig:invg} at the position of the vertical dotted-dashed and dotted-dotted-dashed lines).

This is particularly relevant in the vicinity of the CIR (thick black curves in Fig.~\ref{fig:invg}). The dashed black curve (${B \approx 786.6\,\mathrm{G}-1.5\,\mathrm{G}}$) has a zero crossing between the first and second atom, which results in the exchange coefficients $J_1=0$ and $J_2>0$; the solid black curve (${B \approx 786.6\,\mathrm{G}}$) has a zero crossing at the mean position of the second atom, which results in $J_1=-J_2<0$; finally, the dotted black curve (${B \approx 786.6\,\mathrm{G}+1.5\,\mathrm{G}}$) has a zero crossing between the second and third atom, resulting in $J_1<0$ and $J_2=0$. Therefore, a small change of the magnetic field around the CIR by only $\Delta B \approx 3\,\mathrm{G}$ leads to a strong variation of $J_1/J_2$.

\begin{figure}
\begin{center}
\includegraphics[width = 0.828 \columnwidth]{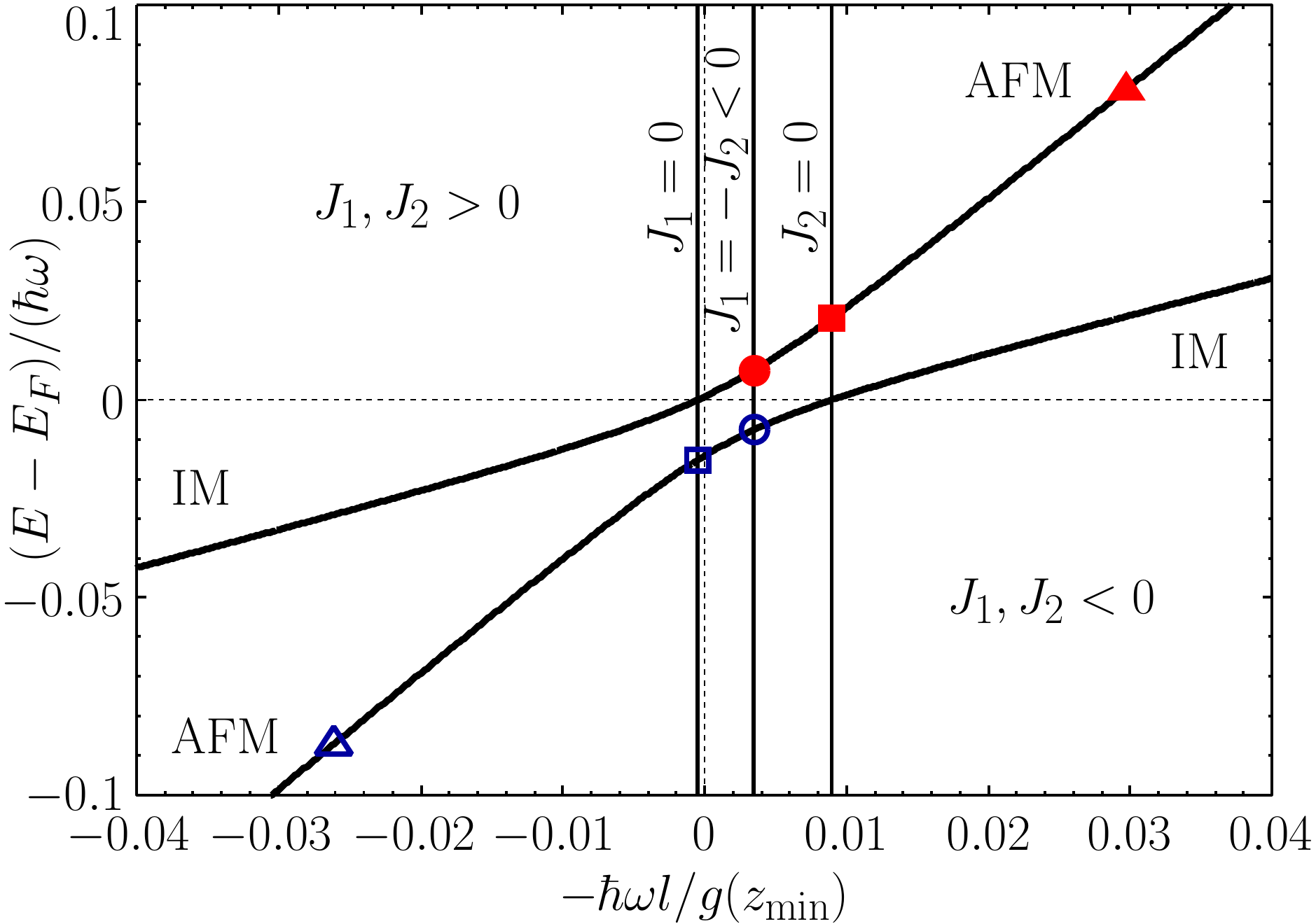}
\caption{Energies of the antiferromagnetic~(AFM) and intermediate~(IM) state as a function of the inverse interaction strength at the trap center $-1/g(z_\mathrm{min})$. The vertical solid lines mark the special cases $J_1 = 0$, $J_1 = -J_2 < 0$, and $J_2 = 0$ (from left to right). The symbols refer to particular states, explained in the text. $\omega$ and $l$ are the frequency and length scale of the harmonic oscillator.}
\label{fig:spectrum}
\end{center}
\end{figure}

This strong change of the ratio $J_1/J_2$ around the CIR causes an avoided crossing between the AFM and IM state. Figure~\ref{fig:spectrum} shows the energies of these states as a function of the inverse interaction strength at the trap center, $-1/g(z_\mathrm{min})$. The avoided crossing is located at $-\hbar \omega l/g(z_\mathrm{min}) \approx 0.0035$ and features a very small splitting, $\Delta E \approx 0.02 \, \hbar \omega$. In the following, we discuss a nonadiabatic sweep across the CIR.

The ramp starts with the AFM state at the open triangle in Fig.~\ref{fig:spectrum} (${J_1 \simeq J_2>0}$). Here, the admixture of the IM state is small, i.e., ${\varphi \simeq 0}$, and the probability that the rightmost spin points downwards is ${P_\downarrow \simeq 16.7 \%}$ [see Eq.~(\ref{eq:spin-orientation})]. By increasing ${-1/g(z_\mathrm{min})}$, the AFM state evolves into the state with the energy marked by the open square (${J_1=0}$, ${J_2>0}$). This state is given by ${| \! \uparrow \rangle \frac{1}{\sqrt{2}} ( | \! \uparrow, \downarrow \rangle - | \! \downarrow, \uparrow \rangle )}$, i.e., the second and third spins form a singlet. The probability that the rightmost spin points downwards is hence ${P_\downarrow = 50 \%}$. Then, the AFM state evolves into the state with the energy marked by the open circle (${J_1=-J_2<0}$). The eigenstate is now given by ${\frac{1}{\sqrt{2}} ( | \mathrm{AFM} \rangle + | \mathrm{IM} \rangle )}$ and ${P_\downarrow = \left( \frac{1}{\sqrt{12}} + \frac{1}{2} \right)^2 \approx 62 \%}$. The nonadiabatic sweep brings the system into the excited state, with the energy marked by the solid circle. This state is given by ${\frac{1}{\sqrt{2}} ( | \mathrm{AFM} \rangle - | \mathrm{IM} \rangle )}$, with ${P_\downarrow = \left( \frac{1}{\sqrt{12}} - \frac{1}{2} \right)^2 \approx 4 \%}$. By further increasing ${-1/g(z_\mathrm{min})}$, the AFM state evolves into the state with the energy at the solid square (${J_1<0}$, ${J_2=0}$), which is given by ${\frac{1}{\sqrt{2}} ( | \! \uparrow, \downarrow \rangle - | \! \downarrow, \uparrow \rangle ) | \! \uparrow \rangle}$, i.e., the singlet is on the left side and ${P_\downarrow = 0 \%}$. Finally, we end up in the AFM state at the solid triangle (${J_1 \simeq J_2<0}$), where again ${\varphi \simeq 0}$ and ${P_\downarrow \simeq 16.7 \%}$ as in the beginning. This discussion clearly shows that a nonadiabatic sweep across the CIR results in the anomalous peak of $P_\downarrow$ observed in Ref.~\cite{Murmann15a}.

We have performed a time-dependent simulation of the experiment, where the system of two spin-up and one spin-down atoms was initially prepared in the noninteracting ground state at a trap depth of ${p = 0.83}$ and a magnetic field of ${B = 523\,\mathrm{G}}$. Then, the magnetic field was ramped up linearly with a ramp speed of ${20\,\mathrm{G} / \mathrm{ms}}$~\cite{Zuern12b} to the final magnetic-field values of ${B = 726\,\mathrm{G} - 1202\,\mathrm{G}}$ around the CIR. The ramp was initially adiabatic, since ${1/\omega \simeq 0.2\,\mathrm{ms}}$, but close to the CIR it was nonadiabatic, since ${1/(0.02 \, \omega) \simeq 10\,\mathrm{ms}}$. Immediately after the $B$-field ramp, the trap depth was linearly lowered to a final value of ${p \approx 0.7 - 0.76}$ within $4\,\mathrm{ms}$ to start the tunneling process. The spilling process lasted ${25 - 200\,\mathrm{ms}}$. The simulation shows that the final state ${| \mathrm{AFM}(t) \rangle}$ after the ramps is not an eigenstate of the time-independent final Hamiltonian. The squared spin overlaps ${|\langle \mathrm{AFM}(t) | f \rangle|^2}$ therefore oscillate around their time averages, since the ramps induce spin dynamics in the spin chain. We therefore choose these time averages and the energy of the nonstationary state to calculate the probability of spin-down tunneling. The result of the calculation is shown in Fig.~\ref{fig:tunneling}. It shows that the improved spin-chain and tunneling model, which includes the $z$ dependence of $1/g$, describes the experiment very well, recovering in particular the anomalous outcoupling peak at the CIR.

\section{Spin chains with FM and AFM exchanges}
\label{sec:unusual-J}

The result of the previous section shows that the transversal confinement may be employed, especially at a CIR, to engineer spatially inhomogeneous spin exchanges in the effective spin chain formed by the strongly interacting spin-$\frac{1}{2}$ fermions. We further illustrate the control of the spin-exchange coefficients by briefly discussing the engineering of a chain that possesses at the same time FM and AFM exchanges. Inspired by Eq.~(\ref{eq:invg}), we choose a position-dependent inverse interaction strength,
\begin{equation}
\mspace{-10mu} \frac{\hbar \omega l}{g(z)} = \sqrt{1 + \alpha \! \left( \! \frac{z-z_0}{l} \! \right)^{\!\!2}} \left[ \beta \sqrt{1 + \alpha \! \left( \! \frac{z-z_0}{l} \! \right)^{\!\!2}} - 1 \right] \!\! ,
\end{equation}
with dimensionless parameters $\alpha$, $\beta$, and $z_0/l$. Here, $\alpha = (l / z_R)^2 \ll 1$, $\beta \propto l_\perp^* / a_\mathrm{3D}(B) \approx 1$ can be tuned by tuning the magnetic field, and $z_0/l$ is the displacement of the harmonic trap from the focus of the laser beam. Another parameter in front of the right-hand side of the above equation, which affects the energy scale, was set to one.

\begin{figure}
\begin{center}
\includegraphics[width = 0.8 \columnwidth]{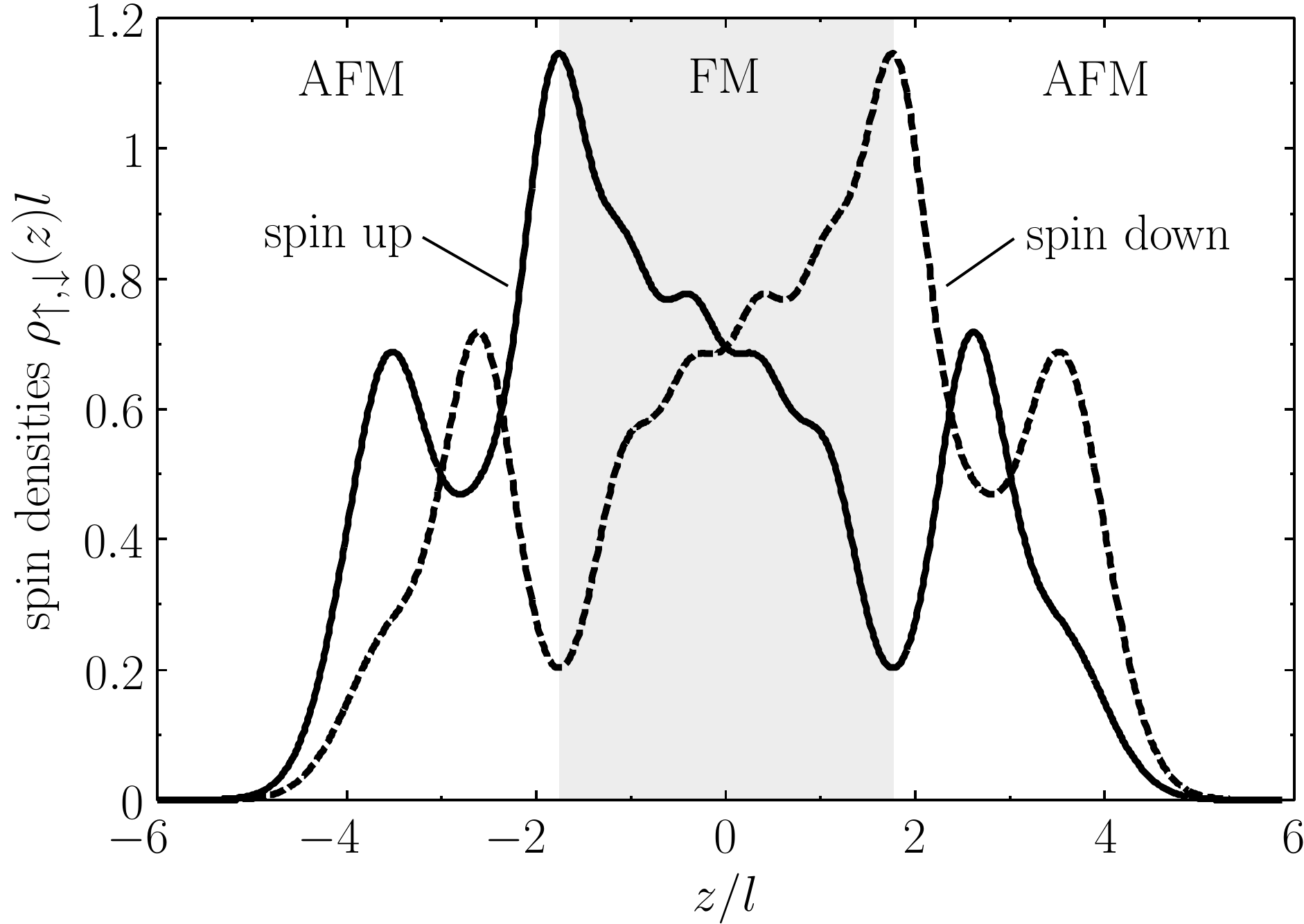}
\caption{Spin densities of five spin-up (solid line) and five spin-down (dashed line) spin-$\frac{1}{2}$ fermions with ferromagnetic~(FM) exchange coefficients ($J_i < 0$) in the center and antiferromagnetic~(AFM) exchange coefficients ($J_i > 0$) at the edges of the spin chain. An additional small linear magnetic-field gradient separates the spin components in the FM phase, but not in the AFM phase. $l$ is the length scale of the harmonic oscillator.}
\label{fig:density}
\end{center}
\end{figure}

By choosing ${\alpha = 0.02}$, ${\beta = 0.98}$, and ${z_0 = 0}$, we obtain a spin chain with FM exchange (${J_i < 0}$) in the center and AFM exchange (${J_i > 0}$) at the edges. The local FM or AFM character may be visualized using a weak linear magnetic-field gradient. The ground-state spin densities~\cite{Deuretzbacher16} are shown in Fig.~\ref{fig:density}. In the FM region, the spin-up and -down components are separated by the magnetic-field gradient, whereas the system exhibits alternating up-down order at the edges of the spin chain, as expected for the AFM region.

\section{Summary and outlook}
\label{sec:conclusions}

We have shown that the exchange coefficients of strongly interacting atoms in a quasi-1D trap may be tuned by varying the transversal confinement along the weak trap axis. This effect is particularly relevant in the immediate vicinity of a CIR. We have shown  that such an inhomogeneity explains the anomalous outcoupling results observed in Ref.~\cite{Murmann15a}. We stress that, although we have illustrated the effect for a particularly simple but experimentally relevant model, similar reasonings may be employed to systems with more atoms and more spin components. Moreover, whereas we have considered the case of a single Gaussian laser beam, more involved transversal confinements may be engineered, including the possibility of changing the transversal confinement in real time. The tunability of the transversal confinement of quasi-1D traps may be hence employed to engineer spin chains with nearly arbitrary (ferro- and antiferromagnetic) exchange coefficients, to prepare other spin states from the antiferromagnetic one using similar avoided crossings as those discussed in this paper, and to initiate and control spin dynamics.

\section*{\uppercase{Acknowledgments}}

We thank D.~Blume and C.~Greene for discussions and comments during the EFB23 conference in Aarhus from August 8th to 12th, 2016, which initiated this work. Moreover, we thank S.~Murmann, G.~Z\"urn, A.~Bergschneider, V.~Klinkhamer, and S.~Jochim for discussions about the experiment. This work was supported by the DFG [Projects No. SA 1031/7-1, No. RTG 1729, and No. CRC 1227 (DQ-mat), Sub-Project A02] and the Cluster of Excellence QUEST.

\bibliographystyle{prsty}

\end{document}